\documentclass[epjCONF]{svjour}
\usepackage{graphics}
\usepackage[varg]{txfonts} 
\usepackage[latin1]{inputenc}
\def\as{\alpha_{\mathrm S}}
\session-title{Hadron Collider Physics Symposium 2011}
\begin{document}
\title{Higgs cross sections: a brief overview}
\author{Massimiliano Grazzini\thanks{On leave of absence from INFN, Sezione di Firenze, Sesto Fiorentino, Florence, Italy}}
\institute{Institute for Theoretical Physics, University of Zurich, CH 8057 Zurich, Switzerland}
\abstract{
I briefly review the status of theoretical predictions for the inclusive production of a Standard Model Higgs boson at hadron colliders. In particular, I focus on the main production channels: gluon-gluon fusion, vector boson fusion and associated production with a vector boson.}
\maketitle
\section{Introduction}
\label{intro}
The search for the Higgs boson and, more generally, the understanding of the origin of electroweak symmetry breaking is one
of the major physics goals of current high-energy colliders.
The Fermilab Tevatron has been shut down in september 2011 having collected
more than 10 fb$^{-1}$ of data.
Combined results with up to 8.6 fb$^{-1}$ integrated luminosity already excluded a Standard Model (SM) Higgs boson in the mass range $156 < m_H < 177$ GeV \cite{CDFandD0:2011aa}: the final
results of the CDF and D0 experiments are expected in 2012.
The CERN LHC, after a successful start of $pp$ collisions in 2009 and 2010, has been operated at a centre-of-mass energy of $7$ TeV in 2011, and data
corresponding to an integrated luminosity of 5.7 fb$^{-1}$ have been accumulated.
These data already allowed the ATLAS \cite{ATLAS} and CMS \cite{CMS} experiments to shrink the
allowed mass range for the SM Higgs boson considerably by essentially excluding
the Higgs bosons in the range ${\cal O}$(130 GeV) $< m_H < {\cal O}$(600 GeV), while observing an excess of Higgs boson candidate events around $m_H= 125$ GeV. More data from the 2012 run are
needed to say whether this is a real Higgs signal or just a statistical fluctuation.

In this contribution I review the current status of theoretical predictions for Higgs boson production at hadron colliders within the SM. I will focus on the main three production channels: gluon--gluon fusion through a heavy quark loop (Sec.~\ref{sec:1}), vector boson fusion (Sec.~\ref{sec:2}), and
associated production with a vector boson (Sec.~\ref{sec:3}).
\section{Gluon fusion}
\label{sec:1}
\begin{figure}
\begin{center}
\resizebox{0.75\columnwidth}{!}{
\includegraphics{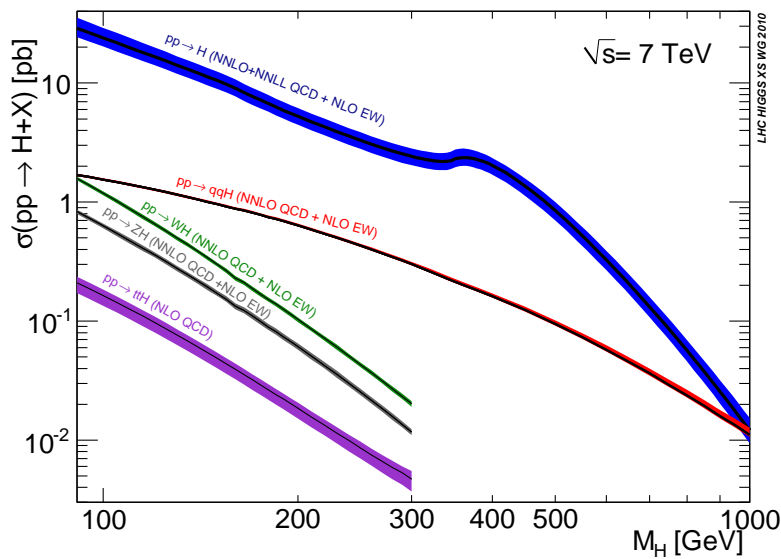}}
\caption{\label{fig:1} Higgs production cross sections at the LHC for $\sqrt{s}=7$ TeV (from Ref.~\cite{Dittmaier:2011ti}).}
\end{center}
\end{figure}

Gluon--gluon fusion through a heavy-quark loop \cite{Georgi:1977gs}
is the main production channel of the SM Higgs boson at hadron colliders.
At the LHC (see Fig.~\ref{fig:1}) the $gg\to H$ cross section is typically
at least one order of magnitude larger than the cross section in the other channels for a wide range of Higgs boson masses. The main contribution comes from the top loop, due to its large Yukawa coupling to the Higgs boson.
The QCD radiative corrections to this process have been computed at next-to-leading order (NLO) both in the large-$m_t$ limit \cite{Dawson:1990zj} and by keeping the exact dependence on the masses of the top and bottom quarks \cite{Djouadi:1991tka,Graudenz:1992pv,Spira:1995rr}.
The impact of NLO correction is very large, of the order of 80-100\% at the LHC, thus casting doubts on the reliability of the perturbative expansion.
The NNLO corrections have been computed in the large-$m_t$ limit \cite{Harlander:2002wh,Anastasiou:2002yz,Ravindran:2003um} and further increase the cross section at the LHC by about 25\%.
Since the completion of the NNLO calculation, the theoretical prediction has been improved in many respect. The logarithmically enhanced contributions due
to multiple soft emissions have been resummed up to next-to-next-to-leading
logarithmic accuracy (NNLL) and the result has been consistently matched to
the fixed order NNLO result \cite{Catani:2003zt}. Soft gluon resummation leads
to an increase of the cross section of about $9\%$ at the LHC ($\sqrt{s}=7$ TeV) and to a slight reduction of scale uncertainties.
Such result \cite{Catani:2003zt} has been used as reference theoretical prediction for few years.
The quantitative impact of soft-gluon resummation is nicely confirmed
by the computation of soft terms at N$^3$LO \cite{Moch:2005ky} (see also \cite{Laenen:2005uz,Idilbi:2005ni,Ravindran:2005vv,Ravindran:2006cg}).

Considerable work has been done also for the computation of EW corrections.
Two-loop EW effects are now known \cite{Djouadi:1994ge,Aglietti:2004nj,Degrassi:2004mx,Actis:2008ts,Actis:2008ug} and their effect strongly depends on the Higgs mass, ranging from $+5\%$ for $m_H=120$ GeV to $-2\%$ for $m_H=300$ GeV \cite{Actis:2008ug}. Mixed QCD-EW effects have been studied in Ref.~\cite{Anastasiou:2008tj}.
EW corrections from real radiation have been studied in Ref.~\cite{Keung:2009bs,Brein:2010xj}: both effects are at the $1\%$ level or smaller.

Quite an amount of work has been devoted to estimate the uncertainties
of the production cross section. The accuracy of the large-$m_t$ approximation
has been studied by computing subleading terms in the large-$m_t$ limit \cite{Marzani:2008az,Harlander:2009bw,Harlander:2009mq,Harlander:2009my,Pak:2009bx,Pak:2009dg}.
Such works have shown that the approximation works remarkably well, to better than $1\%$ for $m_H<300$ GeV. It is fair to say that this was really
a decisive step in having the theoretical prediction for the
$gg\to H$ cross section under good control. In the case of a light Higgs boson produced at the LHC the total theoretical uncertainty is of about $\pm 15-20\%$. We refer the reader to the discussion in Ref.~\cite{Dittmaier:2011ti} for more details.

Various updated calculations on the $gg\to H$ cross section have been presented in the last few years, and we discuss them in turn.
The calculation presented in Ref.~\cite{Anastasiou:2008tj} and refined in Ref.~\cite{Dittmaier:2011ti} starts from the exact NLO QCD calculation (including the dependence on the masses of the top and bottom quarks) and adds the NNLO corrections in the large-$m_t$ limit, and the EW corrections \cite{Actis:2008ug} assuming complete factorization. Mixed QCD-EW effects are evaluated in an effective field theory approach.
It also includes some (small) EW effects from real radiation \cite{Keung:2009bs}. The effect of soft-gluon resummation is mimicked by choosing $\mu_F=\mu_R=m_H/2$ as central values for factorization and renormalization scales.

The calculation of Ref.~\cite{deFlorian:2009hc}, refined in Ref.~\cite{Dittmaier:2011ti},
starts from the exact NLO cross section and includes soft-gluon resummation up to NLL. Then, the top-quark contribution is considered and the NNLL+NNLO corrections \cite{Catani:2003zt} are consistently added in the large-$m_t$ limit.
The result is finally corrected for EW contributions \cite{Actis:2008ug} in the complete factorization scheme. The results of this calculation are available
through an online calculator \cite{hcalc} and are used as reference by the CDF and D0 collaborations at the Tevatron.

The above two calculations show a good agreement over a wide range of Higgs boson masses \cite{Dittmaier:2011ti}.
At the LHC, the reference cross section recommended by the LHC Higgs cross section WG is obtained as a combination of
the results of the above independent calculations \cite{Dittmaier:2011ti}.

Other updated calculations have appeared in the literature.
We first discuss the calculation of Refs.~\cite{Baglio:2010um,Baglio:2010ae}.
As far as the central value is concerned such calculation does not add much to the ones mentioned before. However,
the work of Refs.~\cite{Baglio:2010um,Baglio:2010ae}
presented the first extensive, though extremely conservative, estimate of the various sources of theoretical uncertainties affecting the $gg\to H$ cross section. According to Ref.~\cite{Baglio:2010ae}, the total uncertainty on the $gg\to H$ cross section for a light Higgs boson at the LHC ($\sqrt{s}=7$ TeV) is about $\pm 25\%$.

An independent computation of the inclusive $gg\to H$ cross section
was presented in Ref.~\cite{Ahrens:2010rs}.
Such calculation is based on the NNLO result obtained in the large $m_t$ limit
(the known dependence on top and bottom quark masses up to NLO is not taken into account), corrected with EW effects \cite{Actis:2008ug} and includes the all-order resummation of soft-gluon contributions
according to the formalism presented in Ref.~\cite{Ahrens:2008nc}, with a resummation of the so-called
``$\pi^2$ terms''. This calculation leads to QCD scale uncertainties
of about a factor of three smaller than the calculations discussed above,
and, most likely, not trustable as true perturbative uncertainties.

Recently, a new calculation, implemented in the numerical program {\tt iHixs}
has been presented \cite{Anastasiou:2011pi}. Such calculation includes
essentially the same perturbative contributions of the one of Refs.~\cite{Anastasiou:2008tj,Dittmaier:2011ti} (the additional diagrams considered here give a very small effect). The new important features of {\tt iHixs} are essentially two.
First, it includes finite width effects, allowing the study the Higgs boson lineshape, which is essential
in the searches of a heavy Higgs boson (see also Ref.~\cite{Goria:2011wa}).
Then, it also extends the calculation to models with anomalous Yukawa and electroweak couplings.

\section{Vector boson fusion}
\label{sec:2}
The production of the SM Higgs boson in association with two hard jets
with a large rapidity interval in between, denoted ``vector-boson fusion'' (VBF) is an essential process for the Higgs boson searches at the LHC.
Higgs-boson production in the VBF channel plays an important role also in the determination of Higgs boson couplings.

The production of a Higgs boson + 2 jets receives two contributions at hadron colliders. The first is the genuine VBF process, in which the Higgs boson
is radiated off a vector boson that couples two quark lines.
The hard jets have a strong tendence to be emitted in the forward and backward
directions.
The second contribution is Higgs + 2 jets production through gluon fusion, which represents a background if a measurement of the HWW and HZZ coupling is considered, and interferes with VBF starting from ${\cal O}(\as^2)$.
As a consequence, the genuine VBF process is not completely defined\footnote{VBF interferes already at LO with the associated production with a vector boson, $pp\to HV\to Hjj$, but the effect is at the per mille level \cite{Ciccolini:2007ec}.}.

The NLO QCD corrections to the total VBF rate were computed some time ago
in the so called structure function approach \cite{Han:1992hr}. More recently, the differential cross section at NLO accuracy in QCD has become available \cite{Figy:2003nv}. QCD corrections turn out to be at the level of about 5-10\%.
In Ref.~\cite{Ciccolini:2007jr,Ciccolini:2007ec} combined EW and QCD corrections to VBF have been computed and implemented in a flexible parton level event generator. The impact of EW corrections significantly depends on the Higgs boson mass and for a not too heavy Higgs boson is negative and tends to compensate
the positive effect of QCD corrections.
Other refinements of the vector boson fusion cross section include interference contributions with gluon fusion \cite{Andersen:2006ag,Andersen:2007mp,Bredenstein:2008tm} and gluon induced terms \cite{Harlander:2008xn}. These contributions are well below the percent level.

Approximate NNLO QCD corrections to the total inclusive cross section have been presented in Ref.~\cite{Bolzoni:2010xr,Bolzoni:2011cu}.
The impact of these corrections is extremely small but they further reduce the scale uncertainty down to about $\pm 2\%$. The neglected NNLO diagrams are expected to be both parametrically and kinematically suppressed.

In summary, the VBF channel is under very good theoretical control, since the
theoretical predictions have already a precision comparable to the accuracy
to which the process itself can be defined in perturbation theory.

\section{Associated production with a vector boson}
\label{sec:3}
The production of a Higgs boson in association with a vector boson $V=W^\pm, Z$
is the third important channel at the LHC, as far as the inclusive cross section is concerned (see Fig.~\ref{fig:1}). It is also the most important
search channel in the low mass region at the Tevatron, where the leptonic
decay of the vector boson provides the necessary background rejection.
At the LHC this channel was considered less promising, due to the large backgrounds. In recent years this channel was resurrected by the suggestion \cite{Butterworth:2008iy} to look at events where both the Higgs boson and the vector boson have large transverse momenta.
In such a kinematical region the statistical significance is expected to improve considerably. Needless to say, this channel would provide unique information on the $HWW$ and $HZZ$ couplings.

Up to NLO in QCD perturbation theory the process can be seen as Drell-Yan
production of a vector boson that eventually radiates a Higgs boson.
As such, the QCD corrections up to NLO are identical to those of Drell-Yan \cite{Han:1991ia}.
EW corrections are known and they typically decrease the cross section by about
$5-10\%$ \cite{Ciccolini:2003jy}.
At NNLO QCD corrections are still essentially given by those of Drell-Yan \cite{Hamberg:1990np} and they increase the cross section by about $1-3\%$ at the LHC, and by about $10\%$ at the Tevatron \cite{Brein:2003wg}.
There are, however, additional NNLO diagrams where the Higgs boson is produced
through a heavy quark loop that have to be considered. These diagrams have been recently evaluated in Ref.~\cite{Brein:2011vx}.
At the Tevatron, their effect to $WH$ production is below $1\%$ in the relevant Higgs mass range, while for $ZH$ production, the effect is at the $1-2\%$ level.
At the LHC, the contribution of these terms is typically of the order of $1-3\%$.
In the case of $ZH$ production, since the final state is electrically neutral, there are additional gluon initiated diagrams
that have to be evaluated at NNLO \cite{Brein:2003wg}. Their inclusion is particularly relevant at the LHC,
where the effect ranges from $2\%$ to $6\%$.
Updated predictions for the inclusive $WH$ and $ZH$ cross sections, including
the effect of the additional top-mediated diagrams, are presented
in Ref.~\cite{Brein:2011vx}.

\section{Summary}

In this contribution I have concisely reviewed the current status of
inclusive Higgs production cross sections at the Tevatron and the LHC within the SM, by
focusing on the main three channels: gluon fusion, vector boson fusion and associated production with a vector boson.
I stress that considerable progress has been achieved on the theoretical side
also in the study of {\em differential} and, more generally, {\em fully differential} Higgs production cross sections. More details can be found in \cite{Dittmaier:2012vm}.


\begin{thebibliography}{99}

\bibitem{CDFandD0:2011aa}
  [TEVNPH (Tevatron New Phenomina and Higgs Working Group) and CDF and D0 Collaborations],
  arXiv:1107.5518 [hep-ex].

\bibitem{ATLAS} ATLAS Collaboration, {\em Combination of Higgs Boson Searches with up to 4.9fb$^{-1}$ of pp Collision Data taken at $\sqrt{s}=7$ TeV with the ATLAS Experiment at the LHC}, ATLAS-CONF-2011-163.
\bibitem{CMS} CMS Collaboration, {\em Combination of CMS Searches for a Standard Model Higgs Boson}, CMS-PAS-HIG-11-032.

\bibitem{Georgi:1977gs}
  H.~M.~Georgi, S.~L.~Glashow, M.~E.~Machacek and D.~V.~Nanopoulos,
  Phys.\ Rev.\ Lett.\  {\bf 40} (1978) 692.

\bibitem{Dittmaier:2011ti} 
  S.~Dittmaier {\it et al.}  [LHC Higgs Cross Section Working Group Collaboration],
  arXiv:1101.0593 [hep-ph].


\bibitem{Dawson:1990zj}
  S.~Dawson,
  Nucl.\ Phys.\ B {\bf 359} (1991) 283.

\bibitem{Djouadi:1991tka}
  A.~Djouadi, M.~Spira and P.~M.~Zerwas,
  Phys.\ Lett.\ B {\bf 264} (1991) 440.

\bibitem{Graudenz:1992pv}
  D.~Graudenz, M.~Spira and P.~M.~Zerwas,
  Phys.\ Rev.\ Lett.\  {\bf 70} (1993) 1372.

\bibitem{Spira:1995rr}
  M.~Spira, A.~Djouadi, D.~Graudenz and P.~M.~Zerwas,
  Nucl.\ Phys.\ B {\bf 453} (1995) 17
  [hep-ph/9504378].

\bibitem{Harlander:2002wh}
  R.~V.~Harlander and W.~B.~Kilgore,
  Phys.\ Rev.\ Lett.\  {\bf 88} (2002) 201801
  [hep-ph/0201206].

\bibitem{Anastasiou:2002yz}
  C.~Anastasiou and K.~Melnikov,
  Nucl.\ Phys.\ B {\bf 646} (2002) 220
  [hep-ph/0207004].


\bibitem{Ravindran:2003um}
  V.~Ravindran, J.~Smith and W.~L.~van Neerven,
  Nucl.\ Phys.\ B {\bf 665} (2003) 325
  [hep-ph/0302135].




\bibitem{Catani:2003zt}
  S.~Catani, D.~de Florian, M.~Grazzini and P.~Nason,
  JHEP {\bf 0307} (2003) 028
  [hep-ph/0306211].


\bibitem{Moch:2005ky}
  S.~Moch and A.~Vogt,
  Phys.\ Lett.\ B {\bf 631} (2005) 48
  [hep-ph/0508265].

\bibitem{Laenen:2005uz}
  E.~Laenen and L.~Magnea,
  Phys.\ Lett.\ B {\bf 632} (2006) 270
  [hep-ph/0508284].

\bibitem{Idilbi:2005ni}
  A.~Idilbi, X.~-d.~Ji, J.~-P.~Ma and F.~Yuan,
  Phys.\ Rev.\ D {\bf 73} (2006) 077501
  [hep-ph/0509294].

\bibitem{Ravindran:2005vv}
  V.~Ravindran,
  Nucl.\ Phys.\ B {\bf 746} (2006) 58
  [hep-ph/0512249].

\bibitem{Ravindran:2006cg}
  V.~Ravindran,
  Nucl.\ Phys.\ B {\bf 752} (2006) 173
  [hep-ph/0603041].




\bibitem{Djouadi:1994ge}
  A.~Djouadi and P.~Gambino,
  Phys.\ Rev.\ Lett.\  {\bf 73} (1994) 2528
  [hep-ph/9406432].


\bibitem{Aglietti:2004nj}
  U.~Aglietti, R.~Bonciani, G.~Degrassi and A.~Vicini,
  Phys.\ Lett.\ B {\bf 595} (2004) 432
  [hep-ph/0404071].

\bibitem{Degrassi:2004mx}
  G.~Degrassi and F.~Maltoni,
  Phys.\ Lett.\ B {\bf 600} (2004) 255
  [hep-ph/0407249].



\bibitem{Actis:2008ts}
  S.~Actis, G.~Passarino, C.~Sturm and S.~Uccirati,
  Nucl.\ Phys.\ B {\bf 811} (2009) 182
  [arXiv:0809.3667 [hep-ph]].

\bibitem{Actis:2008ug}
  S.~Actis, G.~Passarino, C.~Sturm and S.~Uccirati,
  Phys.\ Lett.\ B {\bf 670} (2008) 12
  [arXiv:0809.1301 [hep-ph]].


\bibitem{Anastasiou:2008tj}
  C.~Anastasiou, R.~Boughezal and F.~Petriello,
  JHEP {\bf 0904} (2009) 003
  [arXiv:0811.3458 [hep-ph]].


\bibitem{Keung:2009bs}
  W.~-Y.~Keung and F.~J.~Petriello,
  Phys.\ Rev.\ D {\bf 80} (2009) 013007
  [arXiv:0905.2775 [hep-ph]].

\bibitem{Brein:2010xj}
  O.~Brein,
  Phys.\ Rev.\ D {\bf 81} (2010) 093006
  [arXiv:1003.4438 [hep-ph]].



\bibitem{Marzani:2008az}
  S.~Marzani, R.~D.~Ball, V.~Del Duca, S.~Forte and A.~Vicini,
  Nucl.\ Phys.\ B {\bf 800} (2008) 127
  [arXiv:0801.2544 [hep-ph]].

\bibitem{Harlander:2009bw}
  R.~V.~Harlander and K.~J.~Ozeren,
  Phys.\ Lett.\ B {\bf 679} (2009) 467
  [arXiv:0907.2997 [hep-ph]].

\bibitem{Harlander:2009mq}
  R.~V.~Harlander and K.~J.~Ozeren,
  JHEP {\bf 0911} (2009) 088
  [arXiv:0909.3420 [hep-ph]].

\bibitem{Harlander:2009my}
  R.~V.~Harlander, H.~Mantler, S.~Marzani and K.~J.~Ozeren,
  Eur.\ Phys.\ J.\ C {\bf 66} (2010) 359
  [arXiv:0912.2104 [hep-ph]].


\bibitem{Pak:2009bx}
  A.~Pak, M.~Rogal and M.~Steinhauser,
  Phys.\ Lett.\ B {\bf 679} (2009) 473
  [arXiv:0907.2998 [hep-ph]].

\bibitem{Pak:2009dg}
  A.~Pak, M.~Rogal and M.~Steinhauser,
  JHEP {\bf 1002} (2010) 025
  [arXiv:0911.4662 [hep-ph]].

\bibitem{deFlorian:2009hc}
  D.~de Florian and M.~Grazzini,
  Phys.\ Lett.\ B {\bf 674} (2009) 291
  [arXiv:0901.2427 [hep-ph]].


\bibitem{hcalc}
{\tt http://theory.fi.infn.it/grazzini/hcalculators.html}


\bibitem{Baglio:2010um}
  J.~Baglio and A.~Djouadi,
  JHEP {\bf 1010} (2010) 064
  [arXiv:1003.4266 [hep-ph]].

\bibitem{Baglio:2010ae}
  J.~Baglio and A.~Djouadi,
  JHEP {\bf 1103}, 055 (2011)
  [arXiv:1012.0530 [hep-ph]].

\bibitem{Ahrens:2010rs}
  V.~Ahrens, T.~Becher, M.~Neubert and L.~L.~Yang,
  Phys.\ Lett.\  B {\bf 698}, 271 (2011)
  [arXiv:1008.3162 [hep-ph]].

\bibitem{Ahrens:2008nc}
  V.~Ahrens, T.~Becher, M.~Neubert and L.~L.~Yang,
  Eur.\ Phys.\ J.\  C {\bf 62} (2009) 333
  [arXiv:0809.4283 [hep-ph]].


\bibitem{Anastasiou:2011pi}
  C.~Anastasiou, S.~Buehler, F.~Herzog and A.~Lazopoulos,
  JHEP {\bf 1112} (2011) 058
  [arXiv:1107.0683 [hep-ph]].




\bibitem{Goria:2011wa}
  S.~Goria, G.~Passarino and D.~Rosco,
  arXiv:1112.5517 [hep-ph].


\bibitem{Ciccolini:2007ec}
  M.~Ciccolini, A.~Denner and S.~Dittmaier,
  Phys.\ Rev.\ D {\bf 77} (2008) 013002
  [arXiv:0710.4749 [hep-ph]].

\bibitem{Han:1992hr}
  T.~Han, G.~Valencia and S.~Willenbrock,
  Phys.\ Rev.\ Lett.\  {\bf 69} (1992) 3274
  [hep-ph/9206246].


\bibitem{Figy:2003nv}
  T.~Figy, C.~Oleari and D.~Zeppenfeld,
  Phys.\ Rev.\ D {\bf 68} (2003) 073005
  [hep-ph/0306109].


\bibitem{Ciccolini:2007jr}
  M.~Ciccolini, A.~Denner and S.~Dittmaier,
  Phys.\ Rev.\ Lett.\  {\bf 99} (2007) 161803
  [arXiv:0707.0381 [hep-ph]].



\bibitem{Andersen:2006ag}
  J.~R.~Andersen and J.~M.~Smillie,
  Phys.\ Rev.\ D {\bf 75} (2007) 037301
  [hep-ph/0611281].


\bibitem{Andersen:2007mp}
  J.~R.~Andersen, T.~Binoth, G.~Heinrich and J.~M.~Smillie,
  JHEP {\bf 0802} (2008) 057
  [arXiv:0709.3513 [hep-ph]].

\bibitem{Bredenstein:2008tm}
  A.~Bredenstein, K.~Hagiwara and B.~Jager,
  Phys.\ Rev.\ D {\bf 77} (2008) 073004
  [arXiv:0801.4231 [hep-ph]].

\bibitem{Harlander:2008xn}
  R.~V.~Harlander, J.~Vollinga and M.~M.~Weber,
  Phys.\ Rev.\ D {\bf 77} (2008) 053010
  [arXiv:0801.3355 [hep-ph]].

\bibitem{Bolzoni:2010xr}
  P.~Bolzoni, F.~Maltoni, S.~-O.~Moch and M.~Zaro,
  Phys.\ Rev.\ Lett.\  {\bf 105} (2010) 011801
  [arXiv:1003.4451 [hep-ph]].

\bibitem{Bolzoni:2011cu}
  P.~Bolzoni, F.~Maltoni, S.~-O.~Moch and M.~Zaro,
  arXiv:1109.3717 [hep-ph].

\bibitem{Butterworth:2008iy}
  J.~M.~Butterworth, A.~R.~Davison, M.~Rubin and G.~P.~Salam,
  Phys.\ Rev.\ Lett.\  {\bf 100} (2008) 242001
  [arXiv:0802.2470 [hep-ph]].

\bibitem{Han:1991ia}
  T.~Han and S.~Willenbrock,
  Phys.\ Lett.\ B {\bf 273} (1991) 167.

\bibitem{Ciccolini:2003jy}
  M.~L.~Ciccolini, S.~Dittmaier and M.~Kramer,
  Phys.\ Rev.\ D {\bf 68} (2003) 073003
  [hep-ph/0306234].

\bibitem{Hamberg:1990np}
  R.~Hamberg, W.~L.~van Neerven and T.~Matsuura,
  Nucl.\ Phys.\ B {\bf 359} (1991) 343
   [Erratum-ibid.\ B {\bf 644} (2002) 403].

\bibitem{Brein:2003wg}
  O.~Brein, A.~Djouadi and R.~Harlander,
  Phys.\ Lett.\ B {\bf 579} (2004) 149
  [hep-ph/0307206].

\bibitem{Brein:2011vx}
  O.~Brein, R.~Harlander, M.~Wiesemann and T.~Zirke,
  arXiv:1111.0761 [hep-ph].


\bibitem{Dittmaier:2012vm}
  S.~Dittmaier, S.~Dittmaier, C.~Mariotti, G.~Passarino, R.~Tanaka, S.~Alekhin, J.~Alwall and E.~A.~Bagnaschi {\it et al.},
  arXiv:1201.3084 [hep-ph].

\end{thebibliography}
\end{document}